\documentclass[12pt]{article}
\usepackage{graphicx}
\usepackage{fancyhdr}
\renewcommand{\headheight}{27.2pt}

\def\slashchar#1{\setbox0=\hbox{$#1$}           
   \dimen0=\wd0                                 
   \setbox1=\hbox{/} \dimen1=\wd1               
   \ifdim\dimen0>\dimen1                        
      \rlap{\hbox to \dimen0{\hfil/\hfil}}      
      #1                                        
   \else                                        
      \rlap{\hbox to \dimen1{\hfil$#1$\hfil}}   
      /                                         
   \fi}                                         %

\title{\Large \bf Development of a proton Computed Tomography (pCT) scanner at NIU}
\author{S.~A.~Uzunyan$^1$, G.~Blazey$^1$, S.~Boi$^1$, G.~Coutrakon$^1$, \\
                A.~Dyshkant$^1$, B.~Erdelyi$^1$,  A.~Gearhart$^1$,  D.~Hedin$^1$,  \\
                E.~Johnson$^1$, J.~Krider$^1$, V.~Zutshi$^1$,  R.~Ford$^2$, \\
               T.~Fitzpatrick$^2$,  G.~Sellberg$^2$, J.~ E.~Rauch$^2$, M.~Roman$^2$,\\
               P.~Rubinov$^2$, P.~Wilson$^2$,  K.~Lalwani$^3$, M.~Naimuddin$^3$ }
\date{}
\begin{document}
\maketitle
\thispagestyle{fancy}
\begin{center}
\vspace*{-0.3cm}
{\it $^1$~Dept. of Physics, Northern Illinois University, DeKalb,  IL 60115, USA\\
 $^2$~Fermi National Accelerator Laboratory, Batavia, IL 60510, USA\\
 $^3$~Delhi University, 110007, India
}
\end{center}
\vspace{0.3cm}
\begin{center}
{\bf Abstract}\\
\medskip
\parbox[t]{10cm}{\footnotesize
We describe the development of  a proton Computed Tomography (pCT) scanner at Northern Illinois University (NIU) in
collaboration with Fermilab and Delhi University.  
This paper provides an overview of  major components of the scanner and a detailed description of the data acquisition system (DAQ).
%
 }
\end{center}
\newpage
\section{Introduction\label{intro}}
%
Images with protons provide electron density along the proton path in the body of a patient.  
The electron density determines the penetration range for a proton of a certain energy, 
thereby allowing accurate location of the Bragg peak inside a tumor volume.
Proton imaging can provide range uncertainties of about 1\%  compared to 3-4\%  achievable
via traditional X-ray computed tomography, while also inducing a lower dose 
for image production~\cite{pct_uncert}.  
To date a prototype scanner capable of producing images of the required
quality was built at Loma Linda University Medical Center (LLUMC) in 2010~\cite{phaseI_pct}.
The pCT Phase~II scanner constructed at Northern Illinois University (NIU)  is a successor of this device.
It is designed to demonstrate pCT can be used in a clinical environment
and has the  ability to collect data required for 2D or 3D image reconstruction in less than 10 min.
We concentrate here on the data acquisition system.  The detailed description of the scanner hardware components is given in~\cite{brudge_paper}, 
and the image reconstruction hardware and software are described in~\cite{niu_img_reco}.
\section{The scanner design overview \label{design}}
\begin{figure}[ht]
\centering
\includegraphics[scale=.35]{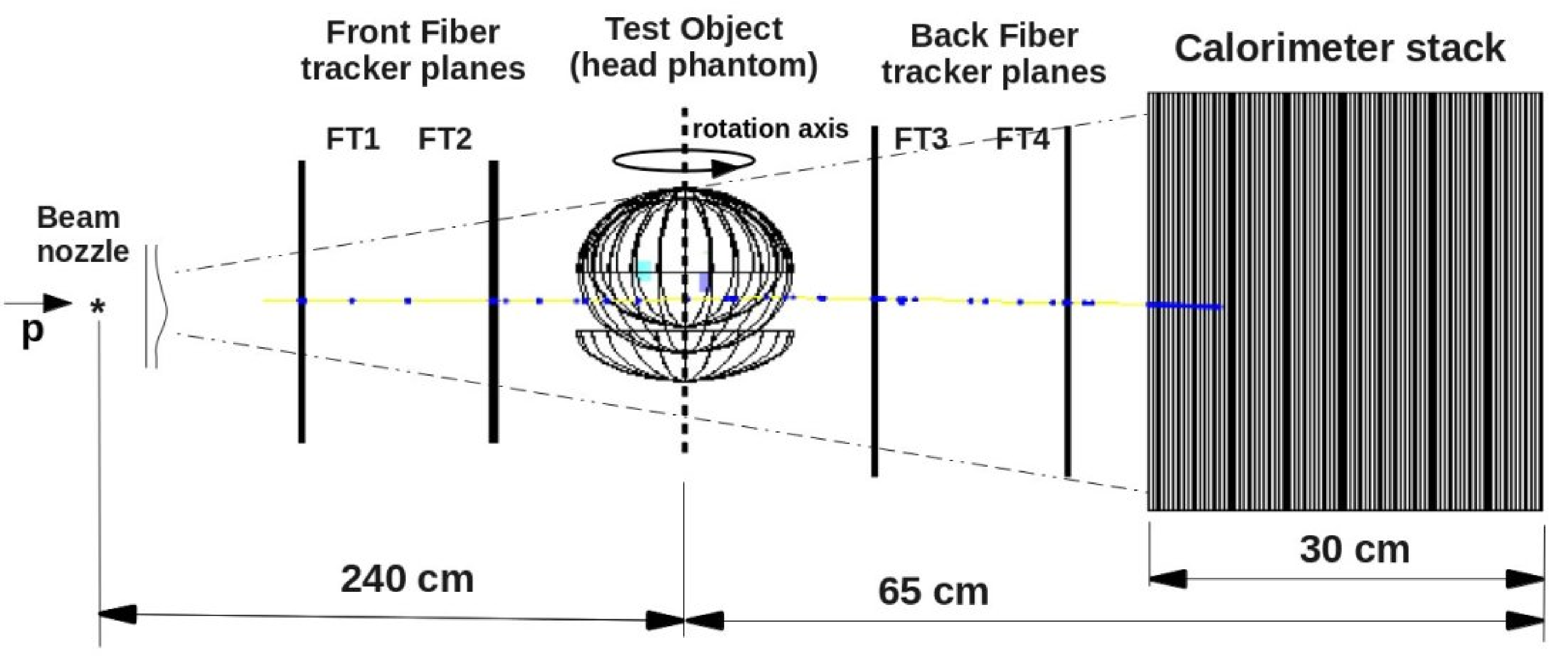}
\caption{\label{fig:pct_Geant} A schematic of the NIU Phase~II pCT detector.}
\end{figure}

The scanner side view is shown in Figure~\ref{fig:pct_Geant},
corresponding to the geometry used for the detector simulation in GEANT~\cite{GEANT}.
The key elements are the fiber tracker (FT) consisting of four X-Y stations (spatial resolution of ~$\sim$1~mm/$\sqrt{12}$)
before and after a rotating Head Phantom, and the range detector, a calorimeter stack consisting of 96, 3.2~mm thick, scintillating tiles. 
The signal readout in both detectors ($\sim$2400 channels) is perfomed with  
CPTA 151-30~\cite{cpta151} silicon photomultipliers (SiPM). For each incident proton the detector measures
the proton track (X,Y) positions in the tracker stations and the residual proton energy deposited in the calorimeter stack.  
The detector acceptance allows scanning of volumes of approximately 24~cm wide and 36~cm high.  The system is designed 
to collect  $\approx2\times10^9$ proton histories for one 3D image of a  human head at  a 
data collection rate of 2~MHz or faster.
\section{The Front-End Data Formats\label{front-end}}
The SiPMs signals from the fiber tracker planes and from the calorimeter stack are collected 
and digitized by the 16 or 32 channel~FPGA-based front-end electronics boards.
The boards send digitized data to the DAQ system through 20  UDP streams (eight are reserved for the fiber tracker 
and 12 for the calorimeter stack) over 1~Gbit/s ethernet connections.  There is no an external trigger:
each board reads out all of its channels if at least one of them has a signal above a threshhold.
The data are shipped in the following formats:\\
$\bullet$ fiber tracker RAW data.  The fibers in the fiber tracker planes are bundled 
in groups of three neighbor fibers. This design allows the incident proton to simultaneously 
hit two adjacent bundles and  thus the front-end reports paired hits:
the local bundle number ($lbn$) of the first bundle in a pair and the state (fired or not) of the $(lbn+1)$ neighbor. 
The timestamp ($ts$) is added to distinguish hits of different proton histories.\newline
$\bullet$ calorimeter stack RAW data.  The scintillator planes in the calorimeter stack are grouped in eight.
For each group the front-end reports: the plane number $LP_{max}$ with the maximum energy deposition, 
the amplitude $A_{max}$ of this maximum energy deposition,
the fractional (to the  $A_{max}$) amplitudes in the remaining seven planes, and the timestamp.\newline
The size of the fiber tracker and calorimeter hits in the described design is  three and six bytes, respectively.
At the readout rate of 2~MHz this requires  6~MB/s transfer rates for the fiber tracker data channels 
(assuming that the level of noise in the fiber tracker planes will be low) and 12~MB/s for the calorimeter data channels.
For  the $2\times10^9$  histories we expect a 208~GB RAW data sample.
\begin{figure}[ht]
\centering
\includegraphics[scale=.33]{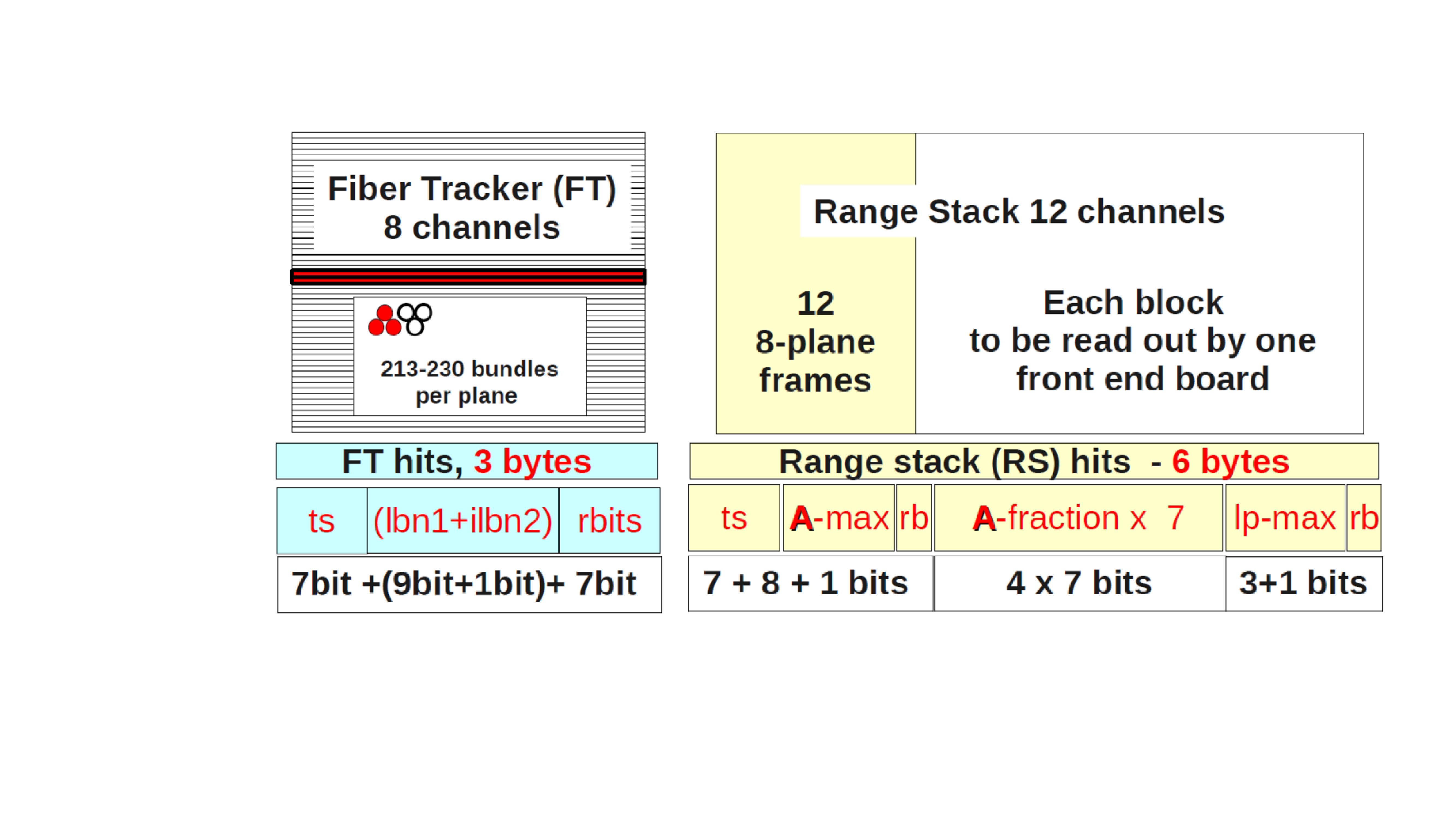}
\leftline{ \hspace{2cm}{\bf (a)} \hfill\hspace{3cm} {\bf (b)} \hfill}
\caption{\label{fig:fe-pade} The bit content of RAW (input) events from the (a) fiber tracker and (b) calorimeter  front-end channels.}
\end{figure}
\section{The DAQ system \label{daq}}
\begin{figure}[ht]
\centering
\includegraphics[scale=0.47]{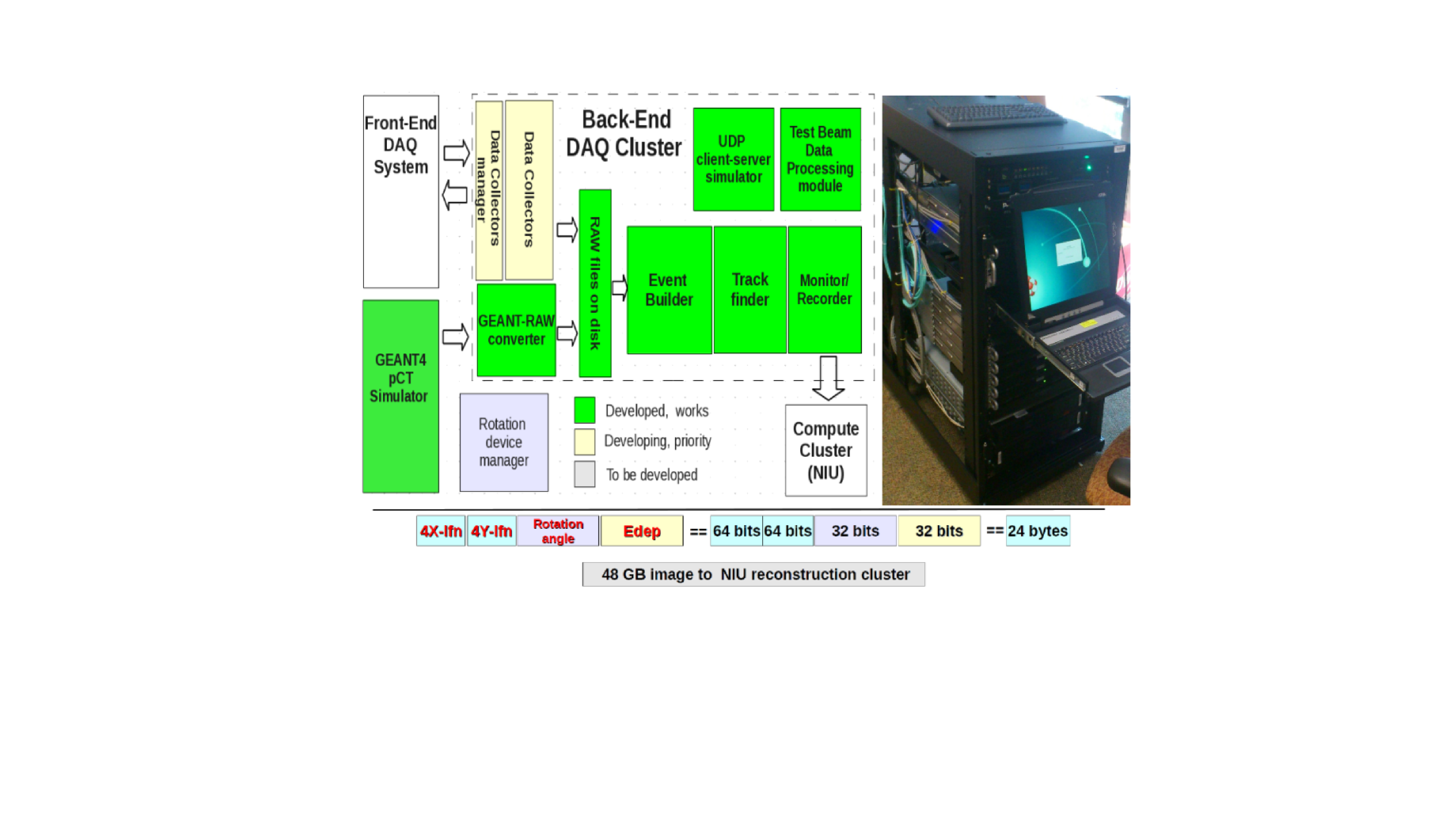}
\caption{\label{fig:pct_daq} A diagram of the DAQ software modules, the assembled DAQ cluster, and the bit content of the output event.}
\end{figure}

The complete DAQ system, shown in Figure~\ref{fig:pct_daq}, was assembled and commissioned in January-March 2013.
The six worker nodes and the head node form a cluster that  provides 24 input
channels to collect  front-end data,  72~CPU cores (running at  2.6~GHz) for the data processing, and 9~TB disk storage space. 
The head node runs cluster management software and is remotely accessible from an operator desktop.
All nodes are interconnected with a 2~Gbit/s internal network. The DAQ software uses the free Scientific Linux 6.2 operating system,
with the event collector and processing modules developed based on the ROOT~\cite{ROOT} data analysis tools.
As tested, this system is capable of accepting data at a rate up to 50~MB/s per input stream with an error rate less than $0.06$\%.  
The maximum amount of RAW data that can be acquired by the cluster during one image
scan is  336~GB (56~GB per worker node).  In the output stream, the DAQ system reconstructs and records each proton track (the eight hits in the fiber planes), the rotation angle of the detector, and the energy deposited in the calorimeter stack. For $2\times10^9$ proton histories 
the 48~GB data file will be stored for subsequent image reconstruction at the NIU Compute Cluster. 
\subsection{Test beam results\label{tbeam}}
\begin{figure}[ht]
\centering
\includegraphics[scale=.30]{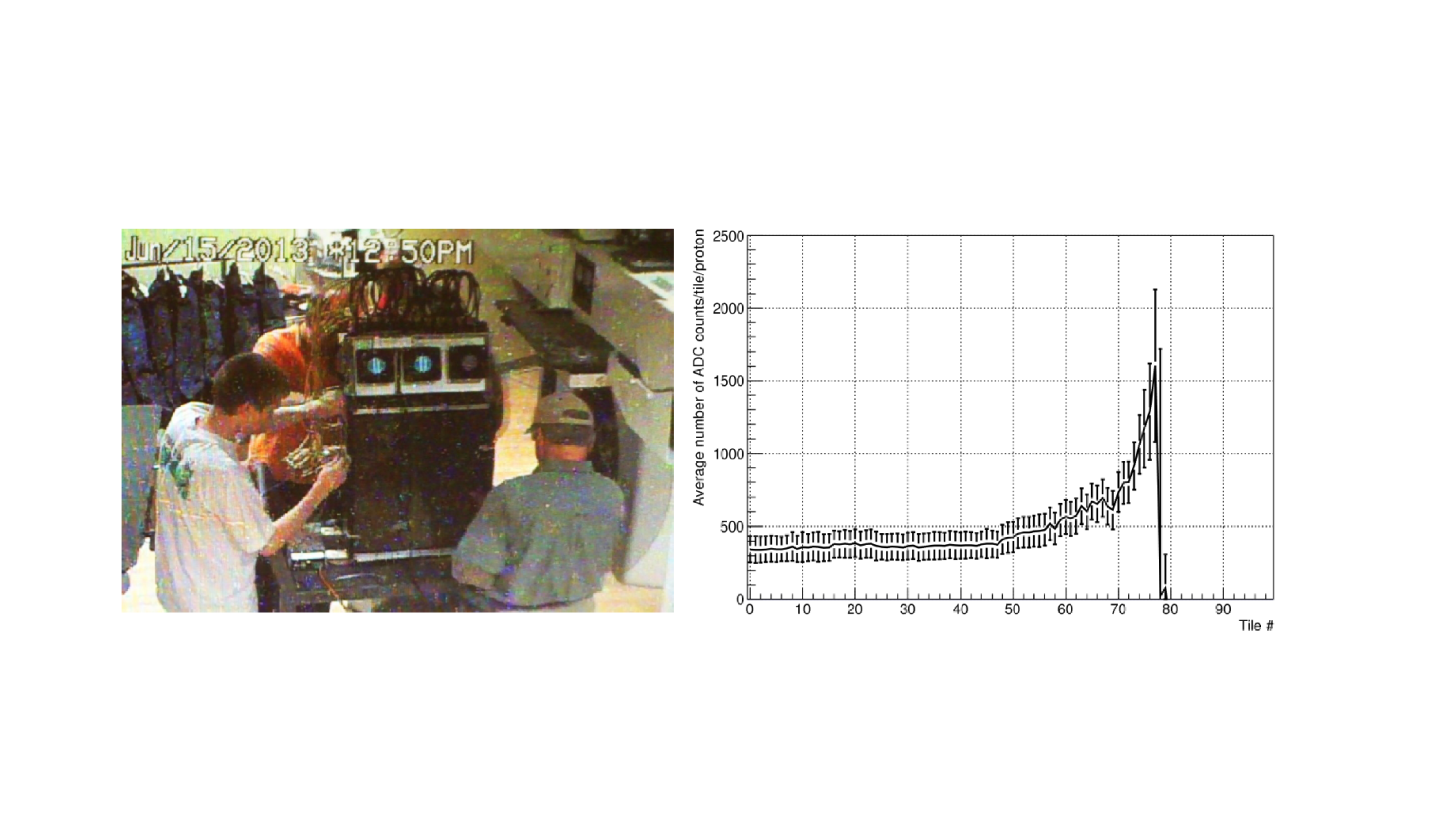}
\leftline{ \hspace{3.5cm}{\bf (a)} \hfill\hspace{3cm} {\bf (b)} \hfill}
\caption{\label{fig:bragg_peak}  a) The assembled calorimeter stack at ProCure Proton center in Warrenville, Illinois;
 b) the average maximum of the calorimeter stack tile signals (in ADC counts) versus tile number collected from ~9000  200~MeV protons. }
\end{figure}

In the Fall of 2012, the DAQ reconstruction software was used for the data taking control and for the data analysis in tests of  
the fiber tracker and calorimeter prototypes at LLUMC.  After assembling the calorimeter this software was again used
for analysis of  tests conducted at the ProCure Proton center in Warrenville, Illinois.  Figure~\ref{fig:bragg_peak} shows
the first results of the Bragg peak measurement for a 200~MeV proton beam. 
\section{The project status}
The major components of the NIU Phase~II pCT scanner (the calorimeter, the fiber tracker and the DAQ system)
were assembled by November 2013 and are being commissioned.
The complete system will be tested in 2014.  The detailed project documentation can be~found~at~\cite{niu_pct_web}. 
%
\section{Acknowledgements}
We thank the staffs at Fermilab and collaborating institutions,
and acknowledge support from the
US Department of Defense.

\end{document}